\documentclass[pre, twocolumn, amsmath, amssymb, superscriptaddress]{revtex4}

\usepackage{graphicx}% Include figure files
\usepackage{dcolumn}% Align table columns on the decimal point
\usepackage{bm}% bold math
\usepackage{braket}
\usepackage{xcolor}

\usepackage{amsmath}

\newcommand{\beq}{\begin{equation}}
\newcommand{\eeq}{\end{equation}}

\usepackage{url}
\usepackage[squaren]{SIunits}

\begin{document}

\title{Zepyros: A webserver to evaluate the shape complementarity of protein-protein interfaces.}

\author{Mattia  Miotto\footnote{\label{corr} For correspondence write to: mattia.miotto@roma1.infn.it}}
\affiliation{Center for Life Nano \& Neuro Science, Istituto Italiano di Tecnologia, Viale Regina Elena 291,  00161, Rome, Italy}

\author{Lorenzo Di Rienzo}
\affiliation{Center for Life Nano \& Neuro Science, Istituto Italiano di Tecnologia, Viale Regina Elena 291,  00161, Rome, Italy}

\author{Leonardo Bo'}
\affiliation{Center for Life Nano \& Neuro Science, Istituto Italiano di Tecnologia, Viale Regina Elena 291,  00161, Rome, Italy}

\author{Giancarlo Ruocco}
\affiliation{Center for Life Nano \& Neuro Science, Istituto Italiano di Tecnologia, Viale Regina Elena 291,  00161, Rome, Italy}
\affiliation{Department of Physics, Sapienza University of Rome, Rome, 00185, Italy.}

\author{Edoardo Milanetti}
\affiliation{Center for Life Nano \& Neuro Science, Istituto Italiano di Tecnologia, Viale Regina Elena 291,  00161, Rome, Italy}
\affiliation{Department of Physics, Sapienza University of Rome, Rome, 00185, Italy.}

\begin{abstract}

Shape complementarity of molecular surfaces at the interfaces is a well-known characteristic of protein-protein binding regions, and it is critical in influencing the stability of the complex. Measuring such complementarity is at the basis of methods for both the prediction of possible interactions and  for the design/optimization of speficic ones. However, only a limited number of tools are currently available to efficiently and rapidly assess it.\\
Here, we introduce Zepyros, a webserver for fast measuring of the shape complementarity between two molecular interfaces of a given protein-protein complex using structural information. Zepyros is implemented as a publicly available tool with a user-friendly interface. Our server can be found at the following  link (all major browser supported):  \url{https://zepyros.bio-groups.com}. 
\end{abstract}

\maketitle

\section{Introduction}

Knowing the stability of protein-protein complexes has important  theoretical implications that are intimately linked to organization of the residue side-chains of the protein partners binding regions~(\cite{hadi2019complex,Desantis2022}), the characterization of which provides information on the nature of the binding of the interacting proteins, and a wide range of potential applications from drug design to the optimization of antibody activity~\cite{norman2020computational,manhart2015protein}. 

Although it is not trivial to define the direct correspondence between the chemical-physical properties of the interfaces and the stability of the protein complex, it is widely known that among the different factors involved in the stabilization of protein complexes, the contribution of the van der Waals forces plays a key role. 
The minimization of the energy contribution of the van der Waals interactions is due to the rearrangement of the side chains of the residues at the interface, thus determining the optimization of the shape complementarity between the molecular surfaces at the interface of interacting proteins.

Indeed, increasing evidence shows that local shape complementary is not only a necessary feature for the identification of binding regions, but also to determine the stability of the resulting complex in contrast with electrostatic compatibility which seems to play a more subtle effect \cite{li2013role,Grassmann2023}, being mostly implicated in molecular recognition~\cite{Spitaleri2018} and structure stability~\cite{Miotto2018, thermo}.

Based on these observations, we previously proposed the innovative approach of representing protein molecular surface in terms of sum of 2D Zernike polynomials, so that comparing the decomposition coefficients between couples of surface patches provides a compact and rapid way to assess shape complementarity (\cite{Milanetti2021}). Our parameter-free method allows to distinguish between binding regions and random portions of the protein molecular surfaces  with an accuracy of $\sim72\%$ and area under the receiver operating characteristic curve (ROC) of 80\% (see Materials and Methods).
Its ability to rapidly quantify local shape complementarity has been used in several applications, ranging from the the understanding/characterization of the properties of different pathological molecular systems~\cite{Grassmann2021, Miotto2022, Piacentini2022, Milanetti2023,  Miotto2023, Monti2024} to the optimization/design of specific protein interactions   
\cite{DiRienzo2021, DeLauro2022, DiRienzo2023, Parisi2024}.

Here, we present Zepyros (ZErnike Polynomials analYsis of pROtein Shapes-Interface Shape Complementary) a free web application that allows for a rapid evaluation of shape complementarity at the interface of bounded protein complexes.

\begin{figure*}
\centering 
\includegraphics[width=\textwidth]{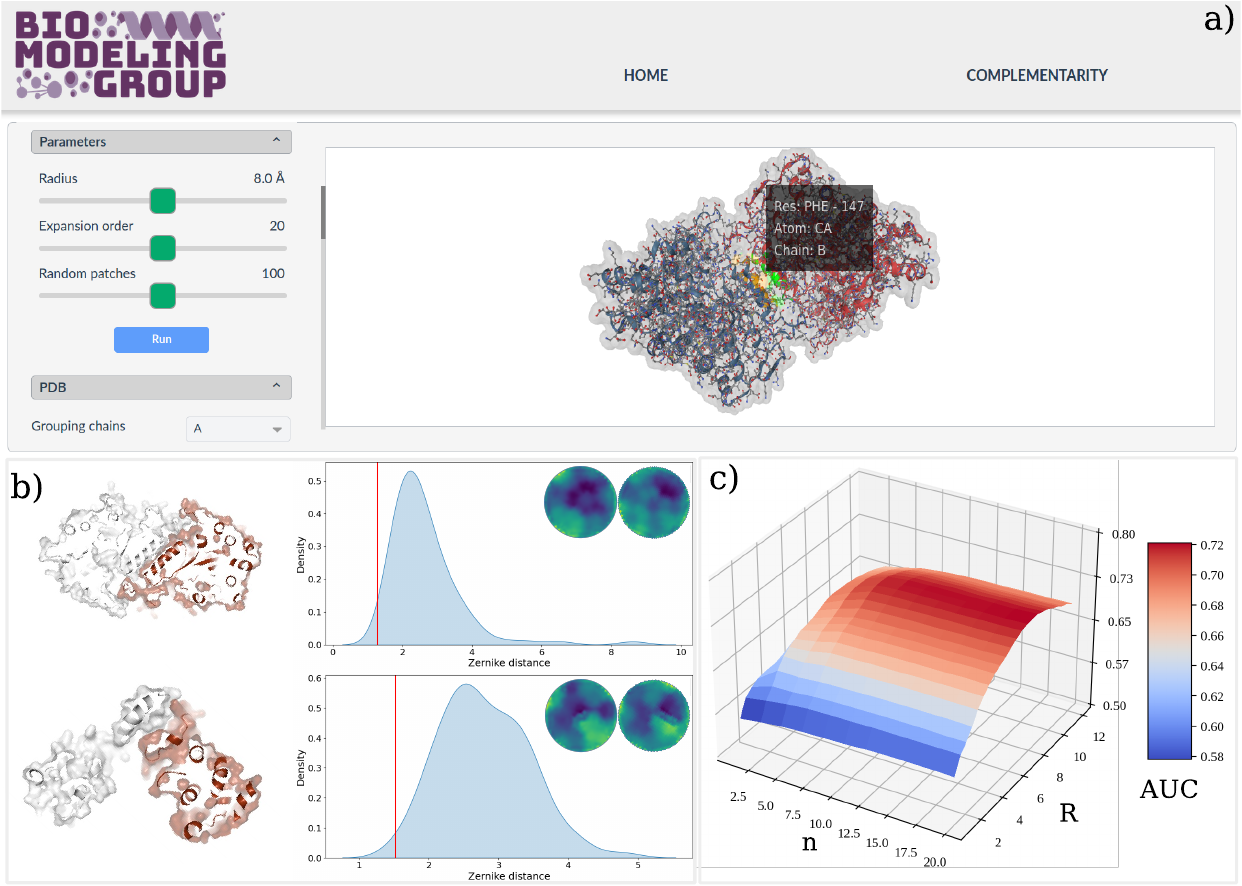}
\caption{\textbf{a)} Input page of the Zepyros server (running examples are provided therein); \textbf{b)}  Example of the results obtained running (top) a homodimer, Ribonucleoside-diphosphate reductase, PDB id: 3O0N; and (bottom) an heterodimer,  the complex of Nuclear pore complex proteins Nup107-Nup133, PDB id: 3CQC. For both cases, a cartoon  representation of the proteins with highlighted molecular surfaces in the binding region are depicted on the left. On the right,  the corresponding Zernike distances between the 2D projections of the binding region is shown as a vertical line together with the distribution of Zernike distances of a set of 100 decoys. 2D projections of the binding patches are represented as color maps above the distributions (see Main text for details). c) Area Under the Curve of the ROC as a function of the order of expansion and patch radius for the 4600 analysed complexes.
}
\label{fig:01}
\end{figure*}
%\enlargethispage{12pt}
%\section{Approach}
%\vspace{-12}

\section{methods}
\subsection{Materials and Methods}
The Zepyros web application allows assessing  the shape complementarity at the binding region of a given protein-protein complex from a starting PDB format (see Figure~\ref{fig:01}a). The tool is based on the following steps:
(i)  the user is required to select to interacting partners, for which the solvent-exposed molecular surfaces are evaluated. Binding regions are identified as the set of residues having surface points closer than $3\angstrom$.
(ii) The identified binding patches are rotated in such a way that the mean normal versor of the template patch is oriented toward the positive z-axis, while the one of the target patch is oriented along the negative z-axis.
(iii) Both patches are projected into a discretized plane so that each pixel of the plane is weigthed according to the distance between the patch points and the origin of a cone containing the patch
(iv) Projected maps ($f(r,\theta)$) are decomposed into the Zernike 2D basis set as 
%\vspace{-9}
\begin{equation}
f(r,\theta) = \sum_{n,m} c_{n}^m Z_n^m(r,\theta)   
%%\vspace{-9}
\end{equation}
where $c_n^m$ are the Zernike complex coefficients and   $Z_n^m$ the Zernike basis polynomials.  
(v) Given two patches, their complementarity can be measured evaluating the Euclidean distance between the squared modulus of the corresponding two sets of coefficients (where the modulus ensure rotational invariance to the resulting descriptors): $ D_z = \sqrt{\sum_{n,m} ( |c_{n}^m| - |{c'}_{n}^m|)^2 }$. 
 %(Figure~\ref{fig:01}f). 

Analysing a large dataset of protein-protein interactions~\cite{Milanetti2021}, we observed that the binding region is characterized on average by a higher-than-random shape complementarity. In particular, comparing the distribution of real Zernike distances with that obtained between couples of random patches, we found that the two distributions are statistically different as measured by the value of AUC of the ROC curve. In Figure~\ref{fig:01}c, we reported the AUC values upon varying the two parameter that the user can set before submitting the job, i.e. the radius of the patches and the order of the expansion.  One can see that while the AUC is always higher than 0.5, the highest scores are obtained for a patch of $\sim 8 \angstrom$ and an order of expansion greater than ten.

 %as shown in  Figure~\ref{fig:01}f.
 % it makes sense to use (a) and (b) but perhaps there is not much space...

\section{Input and output description}

%\subsection{Input}
\textbf{Input}. 
Following the link: \url{https://zepyros.bio-groups.com}, the user reaches the main page (Figure~\ref{fig:01}a), where she/he can upload the PDB file of the protein-protein complex of interest using the upload button. The size of the provided PDB structure must not exceed 10 MB, due to memory issues. 
As sample data, we discuss in the tutorial the case of Ribonucleoside-diphosphate reductase (PDB id: 3O0N). 

%\subsection{Output}

\textbf {Output.} 
The output of a Zepyros run consists of:   
(i)  The Zernike distance between the two interacting patches, providing a measure of the shape complementarity at the interface. While lower the distance the higher the complementarity, we provide a reference frame, comparing the distance with the distribution of distances obtained confronting the target patch with a set of random decoys sampled on the target surface. (ii) The graphical representation of the two real patches projected into the xy plane and confined in the unitary circle  (displayed in Figure~\ref{fig:01}b). (iii) The values of all the computed Zernike coefficients. All data and figures are available in a zip file, which can be downloaded.

%\subsection{Documentation}
\textbf{Documentation}. 
Zepyros has a user-friendly interface that guides how to use the tool and interpret the results. The Documentation page contains a standalone in-depth guide discussing the data submission, as well as of the method and the exploration of the results with case examples.

%\subsection{System requirements}
\textbf{System requirements}. 
The web server requires the most recent version of the following browser with JavaScript enabled: Chrome, Firefox, and Safari.  If your browser connects through a proxy, please, be aware that you might experience a slow upload of the data in the query forms.

\section{Conclusion}
The Zepyros tool is rapid and able to provide results for two average proteins to the user in less than 3 minutes (depending on the selected number of random patches to be evaluated), while for larger complexes  the waiting time is of nearly 5 minutes. The webserver is user-friendly and can be run without any \textit{a priori} knowledge on theoretical or computational biology. We thus believe the measure of shape complemenatity provided by the Zepyros server could be of use in a number of practical applications.
\vspace{-1mm}

\section*{Acknowledgements}
This research was partially funded by grants from ERC-2019-Synergy Grant
(ASTRA, n. 855923); EIC-2022-PathfinderOpen (ivBM-4PAP, n. 101098989);
Project `National Center for Gene Therapy and Drugs based on RNA
Technology' (CN00000041) financed by NextGeneration EU PNRR
MUR—M4C2—Action 1.4—Call `Potenziamento strutture di ricerca e creazione
di campioni nazionali di R\&S' (CUP J33C22001130001).
M.M. acknowledges the CINECA award (SAPPHIR3,  n. HP10BSRYXH) under the ISCRA initiative, for the availability of high performance computing resources and support.

%\bibliographystyle{natbib}
%\bibliographystyle{achemnat}
%\bibliographystyle{plainnat}
%\bibliographystyle{abbrv}
%\bibliographystyle{bioinformatics}
%
%\bibliographystyle{plain}
%
%\bibliographystyle{unsrt}

%\bibliography{mybib}

\end{document}